# A novel picture for charge transport interpretation in epitaxial manganite thin films

P Graziosi[1,2], A Gambardella[1], M Prezioso[1], A Riminucci[1], I Bergenti[1], D Pullini[3], D Busquets-Mataix[2,4] and V A Dediu[1]

[1] CNR - ISMN, Consiglio Nazionale delle Ricerche - Istituto per lo Studio dei Materiali Nanostrutturati, v. Gobetti 101, 40129, Bologna, Italy

[2] Instituto de Tecnología de Materiales, Universitat Politécnica de Valencia, Camino de Vera s/n, 46022, Valencia, Spain

[3] Centro Ricerche Fiat, 10043, Orbassano (TO), Italy

[4] Departamento de Ingeniería Mecánica y de Materiales, Universitat Politécnica de Valencia, Camino de Vera s/n, 46022, Valencia, Spain

E-mail: patrizio.graziosi@gmail.com

**Abstract.** Transport characterizations of epitaxial $La_{0.7}Sr_{0.3}MnO_3$ thin films in the thickness range 5 – 40 nm and 25 – 410 K temperature interval have been accurately collected. We show that taking into account polaronic effects allows to achieve the best ever fitting of the transport curves in the whole temperature range. The Current Carriers Density Collapse picture accurately accounts for the properties variation across the metal-insulator-transition. The electron-phonon coupling parameter $\gamma$ estimations are in a good agreement with theoretical predictions. The results promote a clear and straightforward quantitative description of the manganite films involved in charge transport device applications.

Ferromagnetic manganites are a prototypical example of the so called half metallic materials, i.e. materials with 100% spin polarization at 0 K. Although their possible application in commercial spintronics is prevented by a relatively low Curie temperature ( $T_C \leq 370$ K), manganites represent an ideal laboratory tool to test spin transport in various materials and to search for pioneering device paradigms.[1, 2] Thus, the use of manganites have significantly contributed to the field of organic spintronics, where almost half of the reported devices have $La_{0.7}Sr_{0.3}MnO_3$ (LSMO) as injector.[3] In this context revealing in a most exhaustive and comprehensive way the transport properties of these materials looks both captivating and thought-provoking.

The transport properties of manganites are strongly linked to their ferromagnetism and are generally described in the framework of the so-called double-exchange mechanism: below $T_C$ the exchange interaction between Mn cations through oxygen anions favours ferromagnetism and electron delocalization along the Mn-O-Mn bonds while above $T_C$ thermal disorder disrupts this delocalization leading to a Metal-Insulator Transition (MIT).[4] Nevertheless this basic picture appears to be insufficient to describe the physics of manganites: the strong role of polaron effects in manganites has been pointed out[5] and good experimental evidences for that have been provided by a variety of methods.[5-12] Moreover, manganites were proposed as an example of a *polaronic Fermi liquid*. [6]

Although the presence of polarons in the metallic phase of LSMO (down to 6 K) has been demonstrated by optical conductivity and reflectivity,[7, 8] an explicit evidence of the polaron effects from transport characterizations is still missing.
Indeed, in spite of these convincing optical proves, the dominating trend in literature is to describe the resistivity of manganites via a $T^2+T^5$ behaviour, ignoring thus the polaronic effects (see for example a very recent communication[13]). The few attempts to include polarons in the transport description of the manganites were mainly oversimplified leaving open the issue of real nature of charge carriers. [14, 15]

The standard polaron theory states that at low temperature polarons can form a band and move without thermal activation, carrying their polarization with them.[16] The polaron band narrows with increasing temperature due to the corresponding increase of the polaron effective mass. When the bandwidth is comparable with $k_B T$ the band picture breaks down and the transport becomes thermally activated. This happens at the transition temperature $T_t \approx \frac{\hbar \omega_0}{2 k_B}$, where $\hbar \omega_0$ is the dominant phonon mode coupled to the electron.[17] The polaron band narrowing due to the increase of the polaron effective mass in ionic materials is a known effect and is described by the Sewell model by taking into account the atomicity of the lattice.[17, 18]

As mentioned above most of the attempts tend to promote the description of the transport in LSMO without involving the polaronic effects, though including the electron phonon interaction in the transport equations in four different ways.

(i) A simple combination of electron-phonon interaction $T^5$ term and electron-electron interaction $T^2$ term in $R(T) = R_0 + R_2 T^2 + R_5 T^5$ is often performed neglecting that the $T^5$ dependence of the electron-phonon interaction is just a simplification of the case of a monovalent metal with a spherical Fermi surface and an oversimplified representation of the electron-phonon interaction matrix [19]; indeed the $T^5$ resistivity coefficient appears to vary considerably with temperature and is determined by imposing the $T^5$ dependence in limited temperature ranges even in alkali metals.[19] This description of the electron phonon coupling appears even more inappropriate for LSMO, which has different valence states, carriers concentration and non-spherical Fermi surface.[20]  When applied, [21-25] this approach gave unrealistic values for the $T^5$ coefficient, which is found to be one to two orders of magnitude lower than in metals and, when compared with the electron-electron scattering term, it appears more than 6 orders of magnitude lower than the $T^2$ coefficient while in standard metals it is three to four orders of magnitude lower. [26, 27] These findings are in obvious contradiction with the strong electron-phonon coupling, known to govern the physics of manganites.[5, 11, 15, 28-32] The basic problem is nevertheless the clear failure of this approach to fit the main R(T) features when applied in an extended temperature range, as represented in Figure 1.

(ii) In the second approach the electron-magnon interaction $T^{9/2}$ term[33] substitutes the $T^5$ in the $T^2+T^5$ approach. This does not improve the accuracy of the fit, requiring $T^{9/2}$ coefficients too low to be physically meaningful.[25, 34]

(iii) Even the use of a more generic $T^\alpha$ term, where $\alpha$ is a fitting parameter without a clear link to a scattering mechanism, does not lead to significant improvements.[35]

(iv) Finally, an accurate use of the Bloch-Grüneisen integral was attempted by Varshney and co-workers. [14, 15, 31, 32, 36-39] Their model was found to fit very well the experimental data measured on polycrystalline samples [14, 15, 31, 36-40] while for epitaxial films a good fitting could be achieved only at low temperatures.[32]

As mentioned above the attempts in insert polaronic effects in the transport equations were somehow simplistic. Thus in one of the very few available communications, the polarons have been included by considering the general expression for the mobility $\mu = \frac{e\tau}{m^*}$, where $e$, $m^*$ and $\tau$ are respectively the electronic charge, the effective mass and the scattering time, $\tau$ is assumed to correspond to a polaronic behaviour.[41, 42] A successful fitting was achieved only below 100 K, although the authors used the expression for the scattering time derived by Lang and Firsov for insulating polarons

[43] instead of that for the metallic case.[44] Even the application of the model for this temperature interval was not clearly justified. Indeed, the fitting defined the value $\frac{\hbar\omega_0}{k_B}$ ~80 K, and given that the applicability condition of the model is T ≪ $\frac{\hbar\omega_0}{k_B}$, the analysis should have to be limited to much lower temperatures.

Thus the clear contradiction between transport and optical data sheds doubts on the real role of polarons in these materials and requires an accurate investigation. We show in this paper that transport in high quality LSMO thin films is fully compatible with polaronic picture, which provides an excellent fitting with physically meaningful and realistic parameters.

LSMO epitaxial thin films were deposited by Channel Spark Ablation on $SrTiO_3$ (100) (STO) and $NdGaO_3$ (110) (NGO) substrates.[45] The R(T) measurements were performed in a closed-cycle helium cryostat with four contacts made by silver paste.[46] The polarity inversion method has been used in order to rule out systematic errors.[47] All the measurements starts at 25 K and the heating rate is between 0.2 and 0.3 K/min. The electronic structure of our films, briefly shown elsewhere,[48] is very similar to that known for single crystals, demonstrating the high quality of the films and, as a consequence, the general validity of our findings.

Figures 1 to 3 show detailed experimental data about the resistance as a function of temperature (in log-log scale) collected in 0.1 K steps. While we note that the temperature trend is generally akin to most of the data available in the literature, we concentrate on an explicit quantitative treatment of these curves. We choose to fit the bare measured quantity, i.e. resistance instead of resistivity since we aim to focus on the temperature behaviour of the measured quantity and there is only a geometrical factor. However the resistivity for a typical 40 nm LSMO film on STO is 1.3 mΩ·cm at room temperature and 3.3 mΩ·cm at 340 K, while for 3.6 nm LSMO film on NGO $\rho$ ~ 4.8 mΩ·cm at its MIT (330 K); these value are perfectly in line with the ones known for LSMO films. [49, 50]

We first analyse the metallic parts of the R(T) curves and apply the Matthiessen's rule to the transport properties. In this framework the mobility μ is described as $\frac{1}{\mu} = \sum_i \frac{1}{\mu_i}$, where *i* represents a specific scattering mechanism with its temperature dependence.[51] Furthermore we consider the electrical resistance to be proportional to the inverse of the mobility and assume $\mu = \frac{e\tau}{m^*}$, where $e$, $m^*$ and $\tau$ are respectively the electronic charge, the effective mass and the scattering time.

First we note that R(T) data can be linearized by plotting R vs $T^2$ in a relatively wide range of temperatures, approximately for 25 < T < 60 – 80 K depending on film thickness. It should be noted that such an usually wide interval of validity for a pure $T^2$ dependence is quite characteristic for strongly correlated materials[52-56] although its origin is not yet fully understood.[52] In this paper we take advantage of this to graphically define the fitting coefficients $R_0$ and $R_2$.

In order to understand the transport above 80 K we introduce the electron-phonon interaction. We observe that in strongly polarisable materials such as oxides it is sufficient to consider only the contribution to the electron-phonon interaction of optical phonons.[57] Assuming the Einstein model for lattice vibration we limit the treatment at only one optical phonon $\hbar\omega_0$. Thus the effective mass of the polaron exponentially increases with temperature as $exp\left[g^2 coth\left(\frac{\hbar\omega_0}{2k_BT}\right)\right]$,[17, 18, 58] where $g^2 = \gamma \frac{E_{SP}}{\hbar\omega_0}$ is the zero temperature band-narrowing factor. $E_{SP}$ is the polaronic level shift and represents the amount by which the middle electronic band falls as a result of the potential well created by the lattice deformation, $\gamma$ is the *electron-phonon* interaction parameter. $E_{SP}$ in LSMO is estimated to be about 0.85 eV [29] and $\gamma$ is expected to be less than unity.[57] In the fitting we set $\hbar\omega_0$=425 cm$^{-1}$, which is the most characteristic phonon mode in metallic rhombohedral LSMO. [59-61] In addition to the effective mass approach, we adopt an expression for the scattering time $\frac{1}{\tau} \simeq w\gamma^4 n_\omega(n_\omega + 1)$ in narrow bands, where two phonon scattering is important; $w$ is the bandwidth, $\gamma$ the electron-phonon coupling parameter and $n_\omega = 1/\left(exp\left(\frac{\hbar\omega}{k_BT}\right)-1\right)$ .[62]. The description of the lattice effect on the mobility becomes thus

$$\frac{1}{\mu_{e-ph}} \sim \frac{ew^2\gamma^4 n_\omega(n_\omega+1)}{m_0} \propto exp\left[2g^2 coth\left(\frac{\hbar\omega_0}{2k_BT}\right)\right] n_\omega(n_\omega + 1) . \quad (1)$$

Together with the T$^2$ dependence, empirically found, and considering the Matthiessen's rule, one comes to the final expression for the metallic branch $R_M(T)$ of the total R(T):

$$R_M(T) = R_0 + R_2 T^2 + C\, exp\left[2g^2 coth\left(\frac{\hbar\omega_0}{2k_BT}\right)\right] \frac{exp\left(\frac{\hbar\omega_0}{k_BT}\right)}{\left(exp\left(\frac{\hbar\omega_0}{k_BT}\right)-1\right)^2} . \quad (2)$$

The real fitting parameters are $g^2$ and $C$, the latter representing the proportionality constant between the resistance and the inverse of the electron-phonon dependent mobility; $R_0$ and $R_2$ are obtained graphically and $\hbar\omega_0$ was set as a constant as stated above. We also consider the carriers' density to be constant – we show later that this approximation is consistent.

The fitting results of the $R_M(T)$ with this model are shown in figures 1-3 for 6 different films and the agreement between the experimental data and the model is excellent (green lines). First we note (see table I ) that $C$ is of the same order of magnitude as $R_2$. Next, very similar $\gamma$ values (0.33-0.35) have been obtained for 4 samples out of the 5 deposited on strontium titanate, while it was slightly lower for the neodymium gallate case (0.27). These are reliable values for manganites, close to the ones for cuprates,. about 0.3 .[63] The phonon we have chosen corresponds to a temperature of 600 K, which is about twice the observed MIT temperature, below which the polaron mass model is applicable,[17, 18] and to an energy of 0.05 eV, which is closed to the one used in a recent theory of colossal magnetoresistance (CMR) manganites.[28]

Figure 1 shows the transport data and fits for a 40 nm thick LSMO thin film deposited on NGO. This case is representative for all the samples. It is also clear that polynomial fitting approach (pink line) fails to follow the data above 120 K and that the $T^5$ coefficient is unphysically low, about 9 orders of magnitude lower than the $T^2$ coefficient (we use the $R_5 > 0$ constraint in the numerical fit).

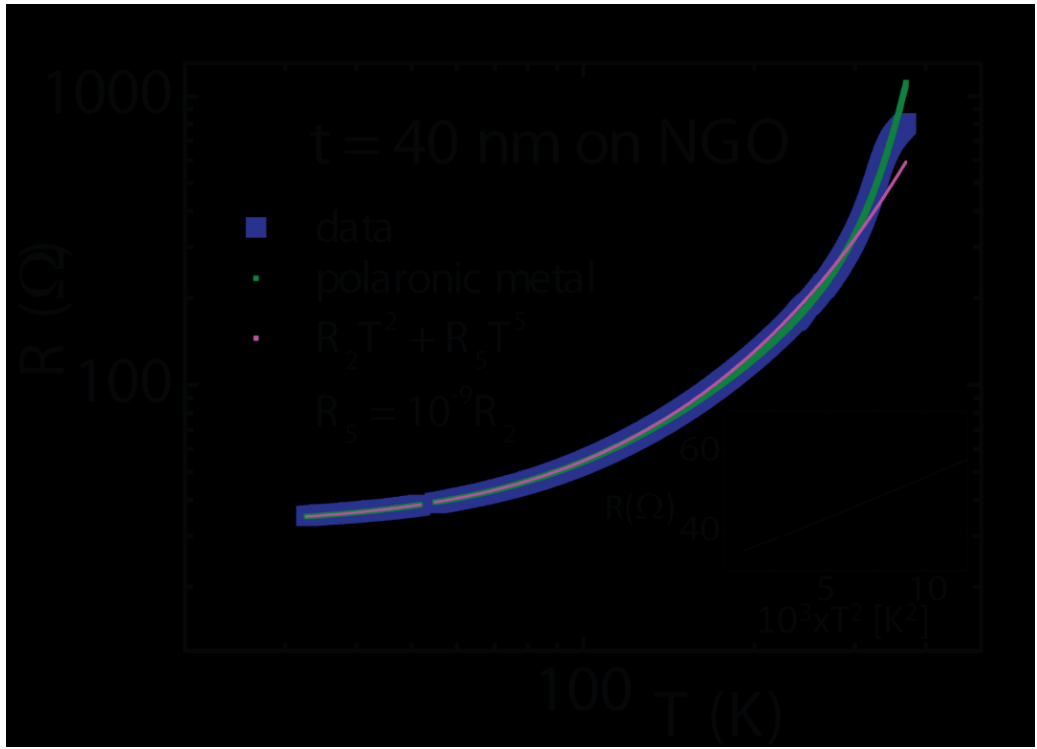

Figure 1: transport data of a 40 nm LSMO/NGO film; the fittings with the polaronic and polynomial models are presented. While the polaronic metal model (eq. (2) )features an excellent agreement to the data, the polynomial model shows poor agreement and unrealistic values of the coefficients. The inset show the linearization of the low T data when plotted versus $T^2$ from which we obtain the $R_0$ and $R_2$ coefficients.

We would like to underline that the term describing the polaronic contribution to the resistance differs from the one used by Varshney and co-workers for the optical phonons[32] only for the effective mass term, confirming that the proper consideration of the temperature effect on the polaron bandwidth was the missing element in their work.

The data above the MIT can be linearized by plotting $\ln(R/T)$ vs $1/T$ (inset to figure 2), indicating standard adiabatic hopping in agreement with both experimental data[25] and polaron theory.[17] In particular this is consistent with the bipolaron theory developed for CMR and enables us to use the Current Carrier Density Collapse (CCDC) picture to describe the MIT.[28, 30, 64] The equation for the insulating branch of R(T) is therefore $R_I(T) = AT exp\left(-\frac{\Delta}{2k_BT}\right)$, where A is a constant and Δ is the bipolaron binding energy. We find Δ ~ 0.1 eV, consistent with the known values for manganites.[21, 30] The overall R(T) is described by

$$R(T) = R_M^{f(T)}(T) \cdot R_I^{1-f(T)}(T),  \quad (3)$$

where $R_M$ and $R_I$ are the fitting equations below and above the MIT and

$$f(T) = \left\{\frac{1}{2}\left(1 - erf\left(\frac{T-T_t}{\Gamma}\right)\right)\right\}^\varsigma \quad (4)$$

Represents the fraction of metallic phase; note that a non-ideality exponent ς is added in respect to the original CCDC model. [28, 30] This way to mix $R_M$ and $R_I$ takes count of the phase separation scenario occurring during the MIT. [65] The whole R(T) is fitted by setting the parameters obtained for the two separate $R_M$(T) and $R_I$(T) branches as constants so that in this final step, the only fitting parameters are those in $f$.

Figures 2 and 3 show the fitting of the insulating phase (red line) for a few films of various thicknesses t deposited on STO and the most relevant fit parameters are reported in table 1. The transition width Γ is relatively broad (25 – 40 K), in agreement with M(T) measurements.[45] The low value of the non-ideality exponent (ς ≈ 1.03) confirms the suitability of the CCDC model. In the CCDC picture the carrier density is almost constant up to a temperature of 0.9·$T_c$ in LSMO,[66] showing that the constant carrier density approximation is realistic in the metallic side, it collapses just when the polaron band picture breaks down.

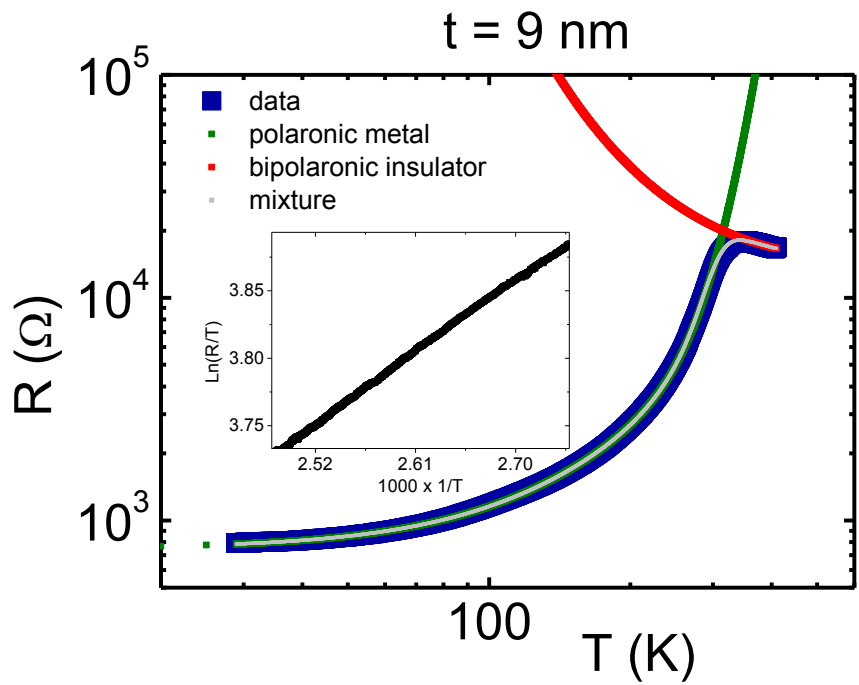

Figure 2: transport data and fittings for a 9 nm LSMO/STO thin film; the metallic side is fitted with eq. (2) while the insulating branch is described by thermal activated adiabatic hopping, which linearizes the data above the MIT (inset). The whole R(T) is fitted by a combination of the twos following the CCDC model , eq. (3).

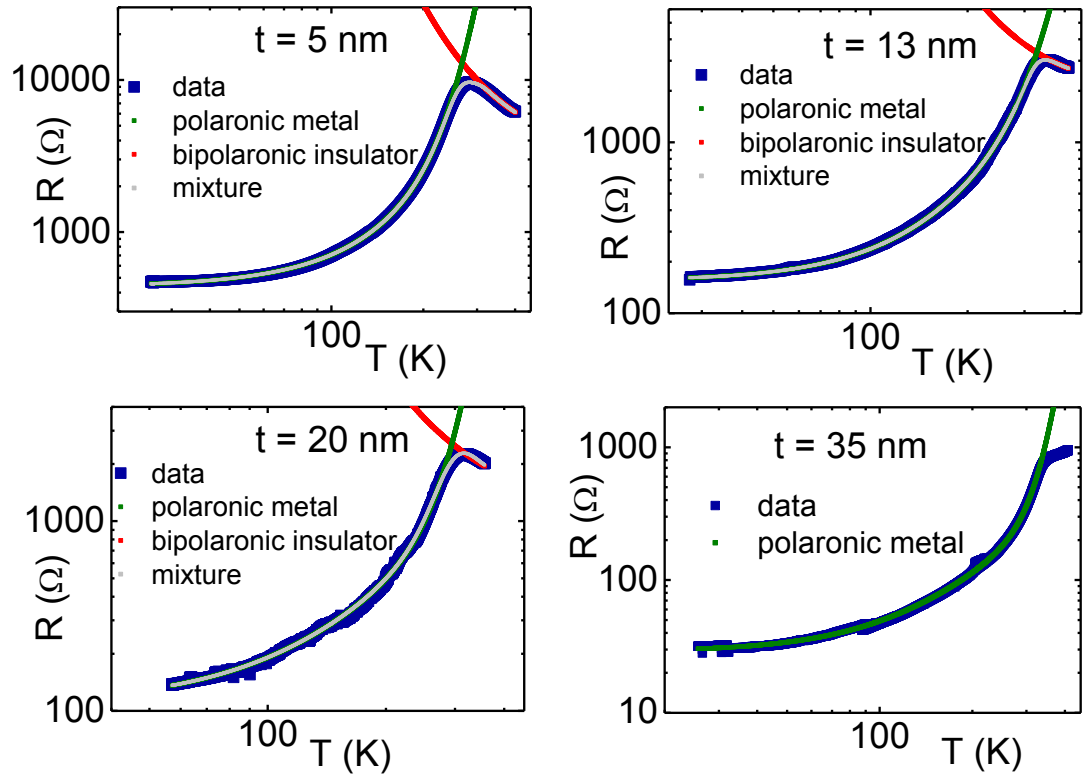

Figure 3: transport data and fittings of films of different thickness t deposited on STO. The metallic side is fitted with eq. (2) while the insulating branch is described by thermal activated adiabatic hopping. The whole R(T) is fitted by a combination of the twos following the CCDC model (eq. (3) ); the 35 nm thick film does not presents the MIT in the investigated range of temperature so only the metallic branch could be fitted.

In summary, we demonstrated that by taking accurately into account the polaron concepts a comprehensive description and an excellent fitting of charge transport characteristics in manganite thin films is achieved both below and above the metal insulator transition. This finding by itself cannot represent a sufficient proof for strong polaronic effects in LSMO. Nevertheless, it adds significant value to optical based proofs promoting polarons as dominant carriers in the ferromagnetic and metallic phase of LSMO and feasibly in other similar manganites.

**Table 1.** Relevant fit parameters

| t (nm) | Substrate | $\gamma$ | $\Gamma$ (K) |
|---|---|---|---|
| 40 | NGO | 0.27 | |
| 5 | STO | 0.33 | 24.2 |
| 9 | STO | 0.33 | 35.6 |
| 13 | STO | 0.19 | 28.1 |
| 20 | STO | 0.35 | 46.2 |
| 35 | STO | 0.35 | |

The authors thank Federico Bona for technical help and Viktor Kabanov for the fruitful discussions. Financial support from the FP7 projects NMP3-LA-2010-246102 (IFOX), NMP-2010-SMALL-4-263104 (HINTS), NMP3-SL-2010-246073 (GRENADA), and Italian government FIRB project n°RBAP117RWN is acknowledged.